# Short Overview of Early Developments of the Hardy Cross Type Methods for Computation of Flow Distribution in Pipe Networks

**Dejan Brkić** [1,2,*,#] **and Pavel Praks** [1,*]

[1] IT4Innovations, VŠB – Technical University of Ostrava, 708 00 Ostrava, Czech Republic
[2] Research and Development Center "Alfatec", 18000 Niš, Serbia
* Correspondence: dejanrgf@tesla.rcub.bg.ac.rs or dejanbrkic0611@gmail.com (D.B.); Pavel.Praks@vsb.cz or Pavel.Praks@gmail.com (P.P.)

Dejan Brkić: https://orcid.org/0000-0002-2502-0601, Pavel Praks https://orcid.org/0000-0002-3913-7800

#Dejan Brkić was on short visit funded by IT4Innovations, VŠB – Technical University of Ostrava

**Abstract:** Hardy Cross originally proposed a method for analysis of flow in networks of conduits or conductors in 1936. His method was the first really useful engineering method in the field of pipe network calculation. Only electrical analogs of hydraulic networks were used before the Hardy Cross method. A problem with flow resistance versus electrical resistance makes these electrical analog methods obsolete. The method by Hardy Cross is taught extensively at faculties, and it remains an important tool for the analysis of looped pipe systems. Engineers today mostly use a modified Hardy Cross method which considers the whole looped network of pipes simultaneously (use of these methods without computers is practically impossible). A method from a Russian practice published during the 1930s, which is similar to the Hardy Cross method, is described, too. Some notes from the work of Hardy Cross are also presented. Finally, an improved version of the Hardy Cross method, which significantly reduces the number of iterations, is presented and discussed. We also tested multi-point iterative methods, which can be used as a substitution for the Newton-Raphson approach used by Hardy Cross, but in this case this approach did not reduce the number of iterations. Although many new models have been developed since the time of Hardy Cross, the main purpose of this paper is to illustrate the very beginning of modelling of gas and water pipe networks and ventilation systems. As a novelty, a new multi-point iterative solver is introduced and compared with the standard Newton-Raphson iterative method.

## 1. Introduction

Hardy Cross solved the problem of distribution of flow in networks of pipes in his article "Analysis of Flow in Networks of Conduits or Conductors" [1] published on 13 November 1936.

Networks of pipes are nonlinear systems since the relation between flow and pressure is not linear. On the contrary, the relation between current and voltage in electrical networks with regular resistors is governed by the linear Ohm's law. Electrical circuits with diodes as well as hydraulic networks are nonlinear systems where resistance depends on current and voltage i.e. on flow and pressure, respectively [2]. Non-linear electrical circuits are electrical circuits containing non-linear components. Nonlinear components can be resistive, capacitive, and inductive.

The distribution of flow in a network of pipes depends on the known inputs and consumptions at all nodes, on the given geometry of pipes, and on network topology. A stable state of flow in a network must satisfy Kirchhoff's laws, which are statements of the conservation of mass and energy. Although there is an indefinite number of flow distributions that satisfy that the conservation of mass is possible in theory, only one distribution from this set also satisfies the conservation of energy for all closed paths formed by pipes in the network. This state is unique for the given network and in- and outflows [3].

Since the relation between flow and pressure is not linear, Hardy Cross used a relation between an increment of flow and an increment of pressure, as this relation is linear for the given quantity of flow. If, however, the increments are fairly large, this linear relation is somewhat in error, as for gas compressible flow. But if the pressure drop in pipes is minor, such as in a municipality network for natural gas distribution, the Hardy Cross method can be used without significant errors [4-6]. Moreover, the Hardy Cross method can also be used for water pipe networks (distring heating [7] and cooling networks [8]) and ventilation systems [9,10] (a related formulation is in Appendix A of this paper).

The Hardy Cross method is an iterative method, i.e. a method using successive corrections [4]. Lobačev and Andrijašev in the 1930s, writing in Russian, offered similar methods [11,12]. Probably because of the language barrier and the political situation in Soviet Russia, Hardy Cross was not aware of Lobačev and Andrijašev's contributions.

Today, engineers use the most improved version of the Hardy Cross method (the $\Delta Q$ method [13]; for $\Delta p$ see [14]), which analyses the whole looped network of pipes simultaneously [15].

As a novel approach presented for the first time here, we tested multi-point iterative methods [16,17] which can be used as a substitution for the Newton-Raphson approach used by Hardy Cross. This approach, however, did not in this case reduce the number of required iterations to reach the final balanced solution.

One example of the pipe network for distribution of gas is analyzed using the original Hardy Cross method [1] in Section 3.1, its related equivalent from Russian literature [11,12] in Section 3.2, the improved version of the Hardy Cross method [15,17] in Section 3.3, and finally the approach which uses multi-point iterative methods instead of the commonly used Newton-Raphson method in Section 3.4.

## 2. Network Piping System; Flow Distribution Calculation

### 2.1. Topology of the Network

The first step in solving a pipe network problem is to make a network map showing pipe diameters, lengths and connections between pipes (nodes). Sources of natural gas supply and consumption rates have to be assigned to nodes. For convenience in locating pipes, code numbers are assigned to each pipe and closed loop of pipes (represented by roman numbers for loops in Figure 1). Pipes on the network periphery are common to one loop and those in the network interior are common to two loops. Figure 1 is an example of a pipe network for distribution of natural gas for consumption in households.

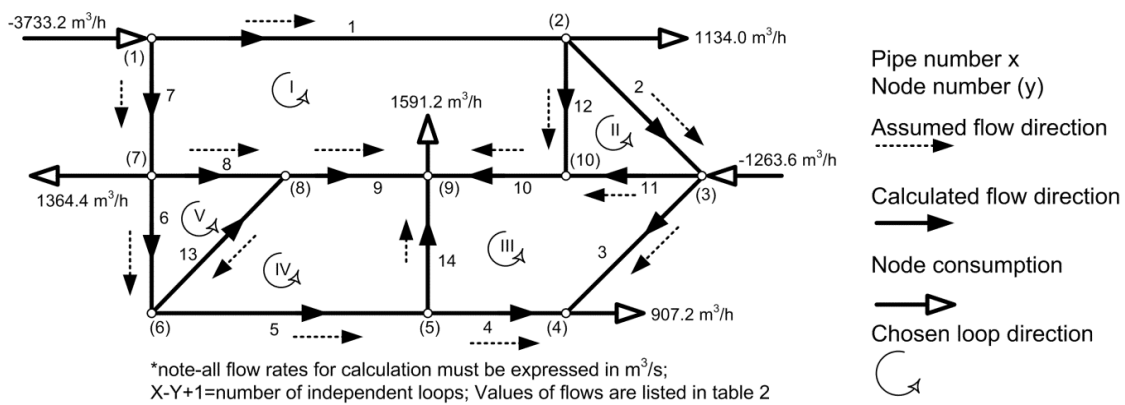

**Figure 1.** The network of pipes for natural gas distribution for domestic consumption

The next step is to write the initial gas flow distribution through pipes in the network. This distribution must be done according to Kirchhoff's first law. The choice of initial flows is not critical, and the criterion should satisfy Kirchhoff's first law for every node in the network [3]. The total gas flow arriving at a node equals the total gas flow that leaves that node. The same conservation law is also valid for the whole network in total (except for gas input and output nodes that cannot be changed during calculations; see consumption nodes in Figure 1). The sum of pseudo-pressure drops along any closed path must be approximately zero for the network to be in balance according to Kirchhoff's second law. In this paper, the flow distribution, which satisfies both of Kirchhoff's laws, will be calculated using the Hardy Cross iterative method.

*2.2. A Hydraulic Model*

The Renouard formula; Eq. (1) best fits a natural gas distribution system built with polyvinyl chloride (PVC) pipes [19,20]. Computed pressure drops are always less than the actual drop since the maximal consumption occurs only during extremely severe winter days [21,22].

$$f = \Delta \tilde{p}^2 = p_1^2 - p_2^2 = 4810 \cdot \frac{\rho_r \cdot L \cdot Q^{1.82}}{D^{4.82}} \tag{1}$$

Where f is a function of pressure, $\varrho_r$ is relative gas density (dimensionless); here $\rho_r = 0.64$, $L$ is the pipe length (m), $D$ is the pipe diameter (m), $Q$ is flow (m³/s), and $p$ is pressure (Pa).

As shown in Appendix A of this paper, other formulas are used in the case of waterworks systems [23,24] and ventilation networks [7].

Regarding the Renouard formula; Eq. (1), one has to be careful since the pressure drop function, $f$, does not relate pressure drop, but actually the difference of the quadratic pressure at the input and the output of the pipe. This means that $\sqrt{\Delta \tilde{p}^2} = \sqrt{p_1^2 - p_2^2}$ is not actually pressure drop despite using the same unit of measurement, i.e. the same unit is used as for pressure (Pa). The parameter $\sqrt{\Delta \tilde{p}^2}$ can be noted as a pseudo pressure drop. In fact the gas is actually compressed, and hence that volume of the gas is decreased, and then such a compressed volume of the gas is conveying with a constant density through the gas distribution pipeline. The operating pressure for a typical distribution gas network is $4x105$Pa abs i.e. $3x105$Pa gauge and accordingly the volume of the gas decreases four times compared to the volume of the gas in normal (or standard) conditions. Pressure in the Renouard formula is for normal (standard) conditions.

The first derivative f' of the Renouard relation; Eq. (2), where the flow is treated as a variable, is used in the Hardy Cross method.

$$f' = \frac{\partial f(Q)}{\partial Q} = 1.82 \cdot 4810 \cdot \frac{\rho_r \cdot L \cdot Q^{0.82}}{D^{4.82}} \tag{2}$$

First assumed gas flow in each pipe is listed in the third column of Table 1. The plus or minus sign preceding flow indicates the direction of flow through the pipe for the particular loop [18,25]. A plus sign denotes counterclockwise flow in the pipe within the loop, while the minus sign, clockwise. The loop direction can be chosen to be clockwise or counterclockwise (in Figure 1 all loops are counterclockwise).

## 3. The Hardy Cross Method; Different Versions

Here will be presented the Hardy Cross method; the original approach in Section 3.1., a version of the Hardy Cross method from Russian practice in Section 3.2., the modified Hardy Cross method in Section 3.3, and finally the method that uses multi-point iterative procedures instead the Newton-Raphson method and which can be implemented in all the aforementioned methods.

*3.1. The Hardy Cross Method; Original Approach*

The pressure drop function for each pipe is listed in Table 1 (for initial flow pattern in the fourth column). The sign in front of the pressure drop function shown in the fourth column is the same as for flow from the observed iteration. The fifth column of Table 1 includes the first derivatives of the pressure drop function, where the flow is treated as a variable. The column of the function of pressure drops is computed algebraically, while the column of the first derivatives is estimated numerically for each loop. Flow correction $\Delta Q$ has to be computed for each loop $x$; Eq. (3).

$$\Delta Q_x = \left(\frac{\sum \pm f}{|f'|}\right)_x = \left(\frac{\sum \pm 4810 \cdot \frac{\rho_r \cdot L \cdot Q^{1.82}}{D^{4.82}}}{\sum \left|1.82 \cdot 4810 \cdot \frac{\rho_r \cdot L \cdot Q^{0.82}}{D^{4.82}}\right|}\right)_x \tag{3}$$

For the network from Figure 1, flow corrections for the first iteration in each loop can be calculated using Eq. (4).

$$\left.\begin{array}{r}(|-f'_1| + |f'_7| + |f'_8| + |f'_9| + |-f'_{10}| + |-f'_{12}|) \cdot \Delta Q_I = -f_1 + f_7 + f_8 + f_9 - f_{10} - f_{12} \\ (|-f'_2| + |-f'_{11}| + |f'_{12}|) \cdot \Delta Q_{II} = -f_2 - f_{11} + f_{12} \\ (|-f'_3| + |f'_4| + |f'_{10}| + |f'_{11}| + |-f'_{14}|) \cdot \Delta Q_{III} = -f_3 + f_4 + f_{10} + f_{11} - f_{14} \\ (|f'_5| + |-f'_9| + |f'_{13}| + |f'_{14}|) \cdot \Delta Q_{IV} = f_5 - f_9 + f_{13} + f_{14} \\ (|f'_6| + |-f'_8| + |f'_{13}|) \cdot \Delta Q_V = f_6 - f_8 + f_{13}\end{array}\right\} \tag{4}$$

In the second iteration, the calculated correction $\Delta Q$ has to be added algebraically to the assumed gas flow (the first initial flow pattern). Further, the calculated correction $\Delta Q$ has to be subtracted algebraically from the gas flow computed in the previous iteration. This means that the algebraic operation for the first correction is the opposite of its sign, i.e. add when the sign is minus, and vice versa. A pipe common to two loops receives two corrections simultaneously. The first correction is from the particular loop under consideration, while the second one is from the adjacent loop, which the observed pipe also belongs to.

Table 1. Procedure for the solution of the flow problem for the network from Figure 1 using the modified Hardy Cross method (first two iterations) – First iteration

| Loop | Pipe | [a]$Q$ | [b]$f = p_1^2 - p_2^2$ | [c]$|f'|$ | [d]$\Delta Q_1$ | [e]$\Delta Q_2$ | [f]$Q_1 = Q$ |
|---|---|---|---|---|---|---|---|
| | | | Iteration 1 | | | | |
| I | 1 | -0.3342 | -144518566.8 | 787025109.2 | -0.0994 | | -0.4336 |
| | 7 | +0.7028 | +859927106.7 | 2226902866.0 | -0.0994 | | +0.6034 |
| | 8 | +0.3056 | +306964191.0 | 1828124435.8 | -0.0994 | -0.0532= | +0.1530 |
| | 9 | +0.2778 | +800657172.4 | 5245486154.8 | -0.0994 | -0.0338= | +0.1446 |
| | 10 | -0.1364 | -241342976.1 | 3220265516.7 | -0.0994 | +0.0142‡ | -0.2217 |
| | 12 | -0.0167 | -6238747.4 | 679911398.4 | -0.0994 | +0.0651‡ | -0.0511 |
| | Σ | | $f_I$=+1575448179.8 | 13987715480.9 | | | |
| II | 2 | -0.0026 | -80628.9 | 56440212.4 | -0.0651 | | -0.0677 |
| | 11 | -0.1198 | -14582531.0 | 221537615.9 | -0.0651 | +0.0142‡ | -0.1707 |
| | 12 | +0.0167 | +6238747.4 | 679911398.4 | -0.0651 | +0.0994 ∓ | +0.0511 |
| | Σ | | $f_{II}$=-8424412.4 | 957889226.7 | | | |
| III | 3 | -0.2338 | -406110098.1 | 3161336093.1 | -0.0142 | | -0.2480 |
| | 4 | +0.0182 | +1530938.1 | 153093808.5 | -0.0142 | | +0.0040 |
| | 10 | +0.1364 | +241342976.1 | 3220265516.7 | -0.0142 | +0.0994 ∓ | +0.2217 |
| | 11 | +0.1198 | +14582531.0 | 221537615.9 | -0.0142 | +0.0651 ∓ | +0.1707 |
| | 14 | -0.0278 | -21840183.8 | 1429824980.5 | -0.0142 | -0.0338 ± | -0.0757 |
| | Σ | | $f_{III}$=-170493836.7 | 8186058014.8 | | | |
| IV | 5 | +0.0460 | +7523646.2 | 297674697.0 | +0.0338 | | +0.0798 |
| | 9 | -0.2778 | -800657172.4 | 5245486154.8 | +0.0338 | +0.0994‡ | -0.1446 |
| | 13 | +0.0278 | +21840183.8 | 1429824980.5 | +0.0338 | -0.0532= | +0.0084 |
| | 14 | +0.0278 | +21840183.8 | 1429824980.5 | +0.0338 | +0.0142 ∓ | +0.0757 |
| | Σ | | $f_{IV}$=-749453158.7 | 8402810812.8 | | | |
| V | 6 | +0.0182 | +3479197.2 | 347919720.0 | +0.0532 | | +0.0714 |
| | 8 | -0.3056 | -306964191.0 | 1828124435.8 | +0.0532 | +0.0994‡ | -0.1530 |
| | 13 | -0.0278 | -21840183.8 | 1429824980.5 | +0.0532 | -0.0338 ± | -0.0084 |
| | Σ | | $f_V$=-325325177.5 | 3605869136.3 | | | |

[a]pipe lengths, diameters and initial flow distribution are shown in Table 2 and Figure 1,

[b]$f$ calculated using the Renouard equation (1),

[c]$f'$ calculated using the first derivative of the Renouard equation (2); flow is variable,

[d]calculated using the matrix equation (10) and entering $\Delta Q$ with the opposite sign (using the original Hardy Cross method for iteration 1: $\Delta Q_I$=+0.1126; $\Delta Q_{II}$=-0.0088; $\Delta Q_{III}$=-0.0208; $\Delta Q_{IV}$=-0.0892; $\Delta Q_V$=-0.0902; using the Lobačev method for iteration 1: $\Delta Q_I$=-0.1041; $\Delta Q_{II}$=-0.0644; $\Delta Q_{III}$=-0.0780; $\Delta Q_{IV}$=+0.1069; $\Delta Q_V$=-0.1824),

[e] $\Delta Q_2$ is $\Delta Q_1$ from the adjacent loop,

[f]the final calculated flow in the first iteration is used for the calculation in the second iteration, etc.,

[g]if $Q$ and $Q_1$ have a different sign, this means that the flow direction is opposite to that in the previous iteration, etc (this occurs with the flow in pipe 13 between iteration 3 and 4).

**Table 1.** *Cont.* – Second iteration

| Loop | Pipe | $Q_1 = Q$ | $f = p_1^2 - p_2^2$ | $|f'|$ | $\Delta Q_1$ | $\Delta Q_2$ | $Q_2 = Q$ |
|---|---|---|---|---|---|---|---|
| | | | Iteration 2 | | | | |
| I | 1 | -0.4336 | -232172997.6 | 974431560.7 | -0.0058 | | -0.4394 |
| | 7 | +0.6034 | +651439280.6 | 1965036192.1 | -0.0058 | | +0.5976 |
| | 8 | +0.1530 | +87112249.4 | 1036457217.8 | -0.0058 | -0.0178= | +0.1294 |
| | 9 | +0.1446 | +243990034.4 | 3070921097.1 | -0.0058 | -0.0098= | +0.1290 |
| | 10 | -0.2217 | -584137977.5 | 4795666298.0 | -0.0058 | +0.0018‡ | -0.2257 |
| | 12 | -0.0511 | -47725420.6 | 1700518680.1 | -0.0058 | -2.1·10⁻⁵ ± | -0.0569 |
| | Σ | | $f_I$=+118505168.7 | 13543031045.9 | | | |
| II | 2 | -0.0677 | -30372941.9 | 816962908.0 | +2.1·10⁻⁵ | | -0.0676 |
| | 11 | -0.1707 | -27780459.9 | 296182372.8 | +2.1·10⁻⁵ | +0.0018‡ | -0.1689 |
| | 12 | +0.0511 | +47725420.6 | 1700518680.1 | +2.1·10⁻⁵ | +0.0058 ∓ | +0.0569 |
| | Σ | | $f_{II}$=-10427981.2 | 2813663960.8 | | | |
| III | 3 | -0.2480 | -451970989.4 | 3317464222.8 | -0.0018 | | -0.2497 |
| | 4 | +0.0040 | +99061.2 | 44589235.4 | -0.0018 | | +0.0023 |
| | 10 | +0.2217 | +584137977.5 | 4795666298.0 | -0.0018 | +0.0058 ∓ | +0.2257 |
| | 11 | +0.1707 | +27780459.9 | 296182372.8 | -0.0018 | -2.1·10⁻⁵= | +0.1689 |
| | 14 | -0.0757 | -135261698.0 | 3251481942.9 | -0.0018 | -0.0098 ± | -0.0873 |
| | Σ | | $f_{III}$=+24784811.3 | 11705384072.0 | | | |
| IV | 5 | +0.0798 | +20483898.1 | 467437803.0 | +0.0098 | | +0.0896 |
| | 9 | -0.1446 | -243990034.4 | 3070921097.1 | +0.0098 | +0.0058‡ | -0.1290 |
| | 13 | +0.0084 | +2454799.0 | 534076127.2 | +0.0098 | -0.0178= | +0.0004 |
| | 14 | +0.0757 | +135261698.0 | 3251481942.9 | +0.0098 | +0.0018 ∓ | +0.0873 |
| | Σ | | $f_{IV}$=-85789639.2 | 7323916970.2 | | | |
| V | 6 | +0.0714 | +41857166.9 | 1067095933.1 | +0.0178 | | +0.0892 |
| | 8 | -0.1530 | -87112249.4 | 1036457217.8 | +0.0178 | +0.0058‡ | -0.1294 |
| | 13 | -0.0084 | -2454799.0 | 534076127.2 | +0.0178 | -0.0098 ± | -0.0004 |
| | Σ | | $f_V$=-47709881.5 | 2637629278.1 | | | |

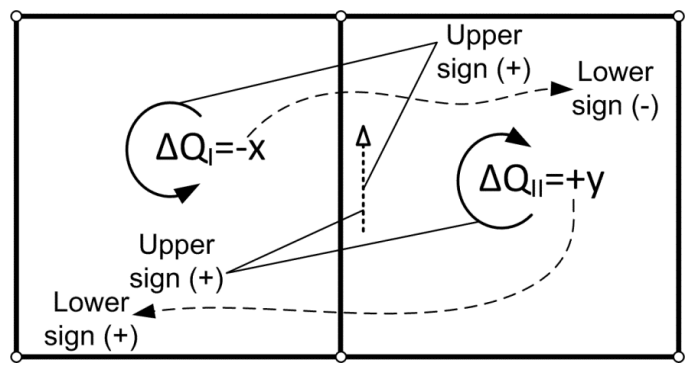

**Figure 2.** Rules for the upper and lower sign (correction from the adjacent loop; second correction)

The upper sign after the second correction in Table 1 is plus if the flow direction in the mutual pipe coincides with the assumed orientation of the adjacent loop, and minus if it does not (Figure 2). The lower sign is the sign in front of correction $\Delta Q$ calculated for the adjacent loop (Figure 2).

Details for signs of corrections can be seen in Brkić [18] and Corfield et al. [25].

The algebraic operation for the second correction should be the opposite of its lower sign when its upper sign is the same as the sign in front of flow $Q$, and as indicated by its lower sign, when its upper sign is opposite to the sign in front of flow $Q$.

The calculation procedure is repeated until the net algebraic sum of pressure functions around each loop is as close to zero as the desired degree of precision demands. This also means that the calculated corrections of flow and the change in calculated flow between two successive iterations is approximately zero. The pipe network is then in approximate balance and the calculation after the Hardy Cross can be terminated.

In the original Hardy Cross method, the corrections for the first iteration are:

$\Delta Q_I = \frac{1575448179.8}{13987715480.9} = +0.1126$,

$\Delta Q_{II} = \frac{-8424412.4}{957889226.7} = -0.0088$,

$\Delta Q_{III} = \frac{-170493836.7}{8186058014.8} = -0.0208$,

$\Delta Q_{IV} = \frac{-749453158.7}{8402810812.8} = -0.0892$, and

$\Delta Q_V = \frac{-325325177.5}{3605869136.3} = -0.0902$.

*3.2. A Version of the Hardy Cross Method from Russian Practice*

As mentioned in the introduction, two Russian authors, Lobačev [11] and Andrijašev [12], proposed a similar method to Hardy Cross [1]. These two methods are also from the 1930s. It is not clear if Hardy Cross had been aware of the contribution of these two authors from Soviet Russia and vice versa, but most probably the answer to this question is no, for both sides. The main difference between the Hardy Cross and Andrijašev methods is that in the method of Andrijašev contours can be defined to include few loops. This strategy only complicates the situation, while the number of required iterations remains unchanged.

Further on the Andrijašev method can be seen from the example in the paper of Brkić [3].

Here the method of Lobačev will be shown in more detail.

In the Hardy Cross method, the influence of adjacent loops is neglected. The Lobačev method takes into consideration this influence; Eq. (5):

$$\left.\begin{array}{r}+(|-f'_1| + |f'_7| + |f'_8| + |f'_9| + |-f'_{10}| + |-f'_{12}|) \cdot \Delta Q_I + |f'_{12}| \cdot \Delta Q_{II} + |f'_{10}| \cdot \Delta Q_{III} + |f'_9| \cdot \Delta Q_{IV} + |f'_8| \cdot \Delta Q_V = -f_1 + f_7 + f_8 + f_9 - f_{10} - f_{12} \\ |f'_{12}| \cdot \Delta Q_I - (|-f'_2| + |-f'_{11}| + |f'_{12}|) \cdot \Delta Q_{II} - |f'_{11}| \cdot \Delta Q_{III} = -f_2 - f_{11} + f_{12} \\ +|f'_{10}| \cdot \Delta Q_I - |f'_{11}| \cdot \Delta Q_{II} - (|-f'_3| + |f'_4| + |f'_{10}| + |f'_{11}| + |-f'_{14}|) \cdot \Delta Q_{III} - |f'_{14}| \cdot \Delta Q_{IV} = -f_3 + f_4 + f_{10} + f_{11} - f_{14} \\ +|f'_9| \cdot \Delta Q_I - |f'_{14}| \cdot \Delta Q_{III} - (|f'_5| + |-f'_9| + |f'_{13}| + |f'_{14}|) \cdot \Delta Q_{IV} - |f'_{13}| \cdot \Delta Q_V = f_5 - f_9 + f_{13} + f_{14} \\ +|f'_8| \cdot \Delta Q_I - |f'_{13}| \cdot \Delta Q_{IV} - (|f'_6| + |-f'_8| + |f'_{13}|) \cdot \Delta Q_V = f_6 - f_8 + f_{13}\end{array}\right\}$$

(5)

In the previous system of equations; Eq. (5), signs in front of terms from the left side of the equals sign have to be determined (this is much more complex than in the Hardy Cross method). So, in the Lobačev method if $(\sum f)_x > 0$ then the sign in front of $(\sum |f'|)_x$ has to be positive, and opposite (for the first iteration this can be seen in Table 1; $f_I = +1575448179.8 > 0$, $f_{II} = -8424412.4 < 0$, $f_{III} = -170493836.7 < 0$, $f_{IV} = -749453158.7 < 0$, $f_V = -325325177.5 < 0$). The sign for other terms (these terms are sufficient in the Hardy Cross method) will be determined using further rules and the scheme from Figure 3.

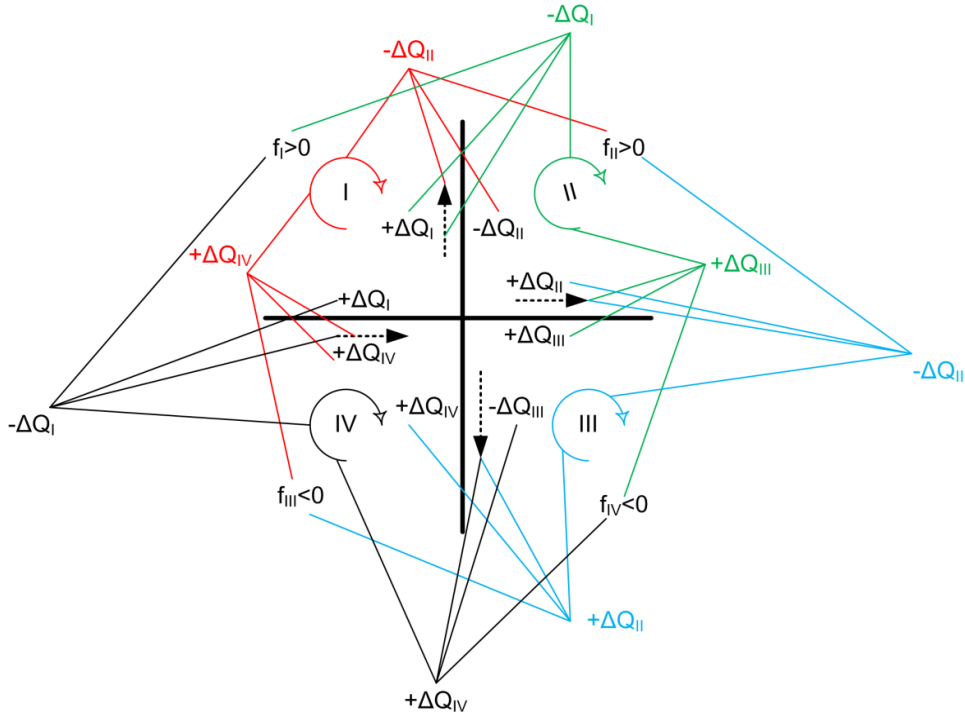

**Figure 3.** Rules for terms from Lobačev equations, which do not exist in the Hardy Cross method

From Figure 3 it can be seen that if $(\sum f)_x > 0$ and if the assumed flow coincides with the loop direction, then the sign of flow in the adjacent pipe is negative and if the flow does not coincide with the loop direction then the sign of flow in the adjacent pipe is positive. Conversely, if $(\sum f)_x < 0$ and if the assumed flow coincides with the loop direction, then the sign of flow in the adjacent pipe is positive and if the flow does not coincide with the loop direction, then the sign of flow in the adjacent pipe is negative. This procedure determines the signs in the front of the flow corrections ($\Delta Q$), which are shown in Figure 3 with black letters (and also in Figure 4 for our pipe network example).

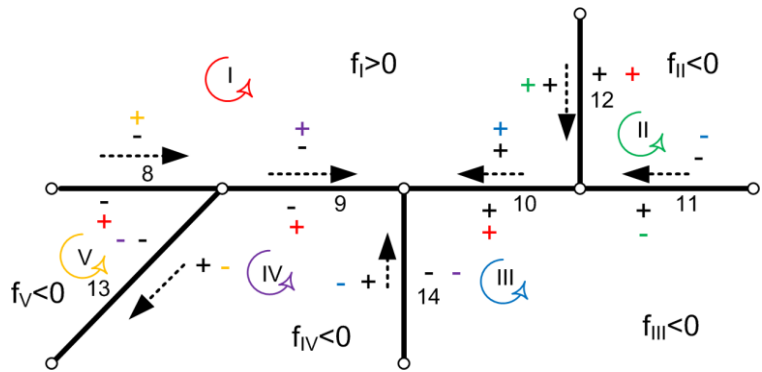

**Figure 4.** Rules for terms from Lobačev equations, which do not exist in the Hardy Cross method applied for the network from Figure 1

If $(\sum f)_x$ from the adjacent loop is positive, while the loop direction and assumed flow do not coincide, the flow correction from the adjacent loop changes its sign, and conversely: if $(\sum f)_x$ from the adjacent loop is positive, while the loop direction and assumed flow coincide, the flow correction from the adjacent loop does not change its sign. If $(\sum f)_x$ from the adjacent loop is negative, while the loop direction and assumed flow do not coincide, the flow correction from the adjacent loop does not change its sign, and conversely: if $(\sum f)_x$ from the adjacent loop is negative, while the loop direction and assumed flow do not coincide, the flow correction from the adjacent loop changes its sign. These four parameters are connected in Figure 3 with the same coloured lines. Flow corrections ($\Delta Q$) shown in Figure 4 with different colours are used with the related signs in Eq. (5). They are chosen in a similar way as explained in the example from Figure 3.

So, instead of the simple equations of the original Hardy Cross method, the system of equations has to be solved in the Lobačev method; Eq. (6).

$$\begin{aligned}
+13987715480.9 \cdot \Delta Q_I + 679911398.4 \cdot \Delta Q_{II} + 3220265516.7 \cdot \Delta Q_{III} + 5245486154.8 \cdot \Delta Q_{IV} + 1828124435.8 \cdot \Delta Q_V &= +1575448179.8 \\
+\underline{679911398.4 \cdot \Delta Q_I} - 957889226.7 \cdot \Delta Q_{II} \underline{-221537615.9 \cdot \Delta Q_{III}} &= -8424412.4 \\
+\underline{3220265516.7 \cdot \Delta Q_I - 221537615.9 \cdot \Delta Q_{II}} - 8186058014.8 \cdot \Delta Q_{III} \underline{-1429824980.5 \cdot \Delta Q_{IV}} &= -170493836.7 \\
+\underline{5245486154.8 \cdot \Delta Q_I - 1429824980.5 \cdot \Delta Q_{III}} - 8402810812.8 \cdot \Delta Q_{IV} \underline{-1429824980.5 \cdot \Delta Q_V} &= -749453158.7 \\
+\underline{1828124435.8 \cdot \Delta Q_I - 1429824980.5 \cdot \Delta Q_{IV}} - 3605869136.3 \cdot \Delta Q_V &= -325325177.5
\end{aligned}$$

(6)

Underlined terms in Eq. (6) do not exist in the Hardy Cross method.

In the Lobačev method, corrections for the first iterations are $\Delta Q_x = \frac{\Delta(\Delta Q_x)}{\Delta}$, where $\Delta$ for the first iteration is; Eq. (7).

$$\Delta = \begin{vmatrix} 13987715480.9 & 679911398.4 & 3220265516.7 & 5245486154.8 & 1828124435.8 \\ 679911398.4 & -957889226.7 & -221537615.9 & 0 & 0 \\ 3220265516.7 & -221537615.9 & -8186058014.8 & -1429824980.5 & 0 \\ 5245486154.8 & 0 & -1429824980.5 & -8402810812.8 & -1429824980.5 \\ 1828124435.8 & 0 & 0 & -1429824980.5 & -3605869136.3 \end{vmatrix} = +3.97 \cdot 10^{+48}$$

(7)

While $\Delta Q_x$ for the first iteration is; Eq. (8).

$$\Delta(\Delta Q_I) = \begin{vmatrix} 1575448179.8 & 679911398.4 & 3220265516.7 & 5245486154.8 & 1828124435.8 \\ -8424412.4 & -957889226.7 & -221537615.9 & 0 & 0 \\ -170493836.7 & -221537615.9 & -8186058014.8 & -1429824980.5 & 0 \\ -749453158.7 & 0 & -1429824980.5 & -8402810812.8 & -1429824980.5 \\ -325325177.5 & 0 & 0 & -1429824980.5 & -3605869136.3 \end{vmatrix} = -4.14 \cdot 10^{+47}$$

(8)

The correction for the first loop in the first iteration is; Eq. (9).

$$\Delta Q_I = \frac{\Delta(\Delta Q_I)}{\Delta} = \frac{-4.14 \cdot 10^{+47}}{+3.97 \cdot 10^{+48}} = -0.1041 \tag{9}$$

Other corrections in the first iteration are $\Delta Q_{II} = -0.0644$, $\Delta Q_{III} = -0.0780$, $\Delta Q_{IV} = +0.1069$ and $\Delta Q_V = -0.1824$.

The Lobačev method is more complex compared to the original Hardy Cross method. But the number of required iterations is not reduced using the Lobačev procedure compared with the original Hardy Cross procedure.

### 3.3. The Modified Hardy Cross Method

The Hardy Cross method can be noted in matrix form. The gas distribution network from Figure 1 has five independent loops; Eq. (10).

$$\begin{bmatrix} \Sigma|f_I'| & 0 & 0 & 0 & 0 \\ 0 & \Sigma|f_{II}'| & 0 & 0 & 0 \\ 0 & 0 & \Sigma|f_{III}'| & 0 & 0 \\ 0 & 0 & 0 & \Sigma|f_{IV}'| & 0 \\ 0 & 0 & 0 & 0 & \Sigma|f_V'| \end{bmatrix} x \begin{bmatrix} \Delta Q_I \\ \Delta Q_{II} \\ \Delta Q_{III} \\ \Delta Q_{IV} \\ \Delta Q_V \end{bmatrix} = \begin{bmatrix} \Sigma \pm f_I \\ \Sigma \pm f_{II} \\ \Sigma \pm f_{III} \\ \Sigma \pm f_{IV} \\ \Sigma \pm f_V \end{bmatrix} \tag{10}$$

Eq. (4) provides for each particular loop in the network the same corrections as Eq. (10) using matrix calculation. Epp and Fowler [15] improved the original Hardy Cross method [1] by replacing some of the zeroes in the non-diagonal terms of Eq. (10). For example, if pipe 8 is mutual for loop I and V, the first derivative of the pressure drop function for the observed pipe, where flow treated as a variable, will be put with a negative sign in the first column and the fifth row, and also in the fifth column and the first row; Eq. (11).

$$\begin{bmatrix} \sum|f'_I| & -f'_{12} & -f'_{10} & -f'_9 & -f'_8 \\ -f'_{12} & \sum|f'_{II}| & -f'_{11} & 0 & 0 \\ -f'_{10} & -f'_{11} & \sum|f'_{III}| & -f'_{14} & 0 \\ -f'_9 & 0 & -f'_{14} & \sum|f'_{IV}| & -f'_{13} \\ -f'_8 & 0 & 0 & -f'_{13} & \sum|f'_V| \end{bmatrix} x \begin{bmatrix} \Delta Q_I \\ \Delta Q_{II} \\ \Delta Q_{III} \\ \Delta Q_{IV} \\ \Delta Q_V \end{bmatrix} = \begin{bmatrix} \sum \pm f_I \\ \sum \pm f_{II} \\ \sum \pm f_{III} \\ \sum \pm f_{IV} \\ \sum \pm f_V \end{bmatrix} \quad (11)$$

In the modified Hardy Cross method, corrections for the first iterations are (12); where solutions are listed in Table 1.

$$\begin{bmatrix} +13987715480.9 & -679911398.4 & -3220265516.7 & -5245486154.8 & -1828124435.8 \\ -679911398.4 & +957889226.7 & -221537615.9 & 0 & 0 \\ -3220265516.7 & -221537615.9 & +8186058014.8 & -1429824980.5 & 0 \\ -5245486154.8 & 0 & -1429824980.5 & +8402810812.8 & -1429824980.5 \\ -1828124435.8 & 0 & 0 & -1429824980.5 & +3605869136.3 \end{bmatrix} x \begin{bmatrix} \Delta Q_I \\ \Delta Q_{II} \\ \Delta Q_{III} \\ \Delta Q_{IV} \\ \Delta Q_V \end{bmatrix} = \begin{bmatrix} +1575448179.8 \\ -8424412.4 \\ -170493836.7 \\ -749453158.7 \\ -325325177.5 \end{bmatrix}$$
(12)

This procedure significantly reduces the number of iterations required for the solution of the problem (Figure 5).

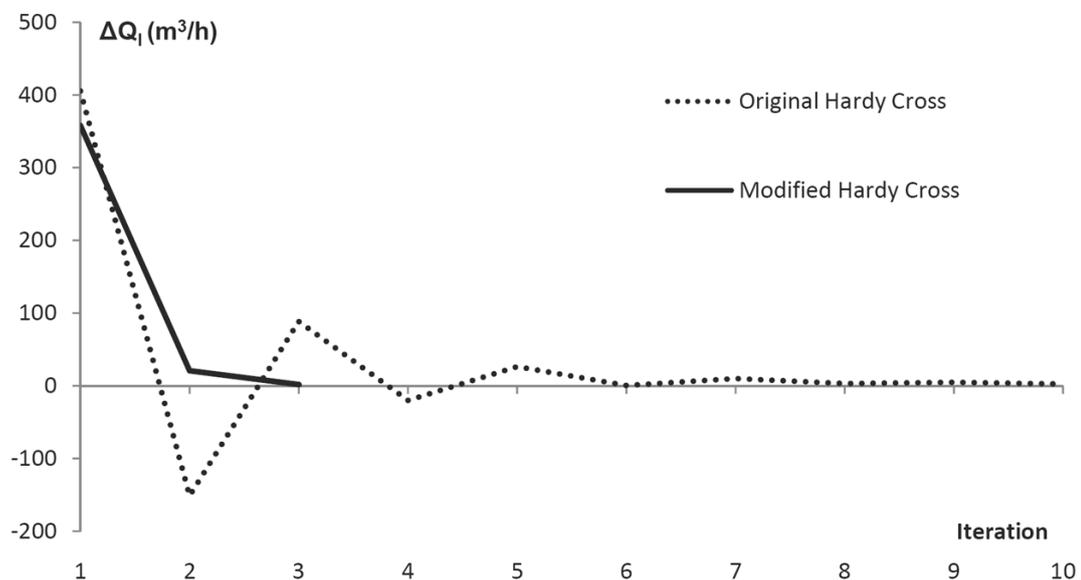

**Figure 5.** The number of required iterations for the solution using the original vs. the improved Hardy Cross method

The first two iterations for the network from Figure 1 are shown in Table 1. Pipe diameters and lengths, as well as the first, assumed, and the final calculated flow distributions for the network in balance are shown in Table 2.

The gas velocity in the network is small (can be up to 10-15 m/s). The network can be the subject of diameter optimization (as in [4]), which can also be done by using the Hardy Cross method (diameter correction $\Delta D$ should be calculated for known and locked flow, where the first derivative of the Renouard function has to be calculated for diameter as a variable). The network should stay unchanged, even if planned gas consumption on nodes 5, 6, 8 and 10 increases, as pipes 4 and 13 will be useful thanks to increased gas flow.

Similar examples, but for water flow, can be seen in [26]. Optimization of pipe diameters in a water distributive pipe network using the same approach can be seen in [6].

Table 2. Pipe diameters and lengths, flows, and velocities of gas within pipes

| [a]Pipe number | Diameter (m) | Length (m) | [b]Assumed flows (m³/h) | [c]Calculated flows (m³/h) | Gas velocity (m/s) |
|---|---|---|---|---|---|
| 1 | 0.305 | 1127.8 | 1203.2 | 1583.6 | 1.5 |
| 2 | 0.203 | 609.6 | 9.2 | 245.2 | 0.5 |
| 3 | 0.203 | 853.4 | 841.6 | 899.7 | 1.9 |
| 4 | 0.203 | 335.3 | 65.6 | 7.5 | 0.01 |
| 5 | 0.203 | 304.8 | 165.6 | 320.2 | 0.7 |
| 6 | 0.203 | 762.0 | 65.6 | 322.7 | 0.7 |
| 7 | 0.203 | 243.8 | 2530.0 | 2149.6 | 4.6 |
| 8 | 0.203 | 396.2 | 1100.0 | 462.4 | 1.0 |
| 9 | 0.152 | 304.8 | 1000.0 | 465.0 | 1.8 |
| 10 | 0.152 | 335.3 | 491.2 | 813.5 | 3.1 |
| 11 | 0.254 | 304.8 | 431.2 | 609.1 | 0.8 |
| 12 | 0.152 | 396.2 | 60.0 | 204.8 | 0.8 |
| 13 | 0.152 | 548.6 | 100.0 | [d]-2.6 | -0.009 |
| 14 | 0.152 | 548.6 | 100.0 | 312.7 | 1.2 |

[a]network from Figure 1 (flows are for normal pressure conditions; real pressure in the network is $4x10^5$ Pa abs i.e. $3x10^5$ Pa)

[b]chosen to satisfy Kirchhoff's first law for all nodes (dash arrows in Figure 1)

[c]calculated to satisfy Kirchhoff's first law for all nodes and Kirchhoff's second law for all closed path formed by pipes (full errors in Figure 1)

[d]the minus sign means that the direction of flow is opposite to the initial pattern for assumed flows

*3.4. The Multi-Point Iterative Hardy Cross method*

The here described multipoint method can substitute the Newton-Raphson iterative procedure used in all the above described methods. Recently, we successfully used the here presented multipoint method for acceleration of the iterative solution of the Colebrook equation for flow friction modelling [16,17]. On the contrary, for the gas network example of Figure 1, the multipoint method requires the same number of iterations as the original Newton-Raphson procedure.

For the test, we used the three-point method from Džunić et al. [27]. Flow corrections $\Delta Q_I$ from Eqs. (10) and (11) from the first loop I should be calculated using the three-point procedure; Eq. (13):

$$\begin{cases} \Delta Q_I = -\frac{(\Sigma \pm f_I)_i}{(\Sigma |f_I'|)_i}, \text{where } i = 1; \\ \Delta Q_I = -\frac{(\Sigma \pm f_I)_i}{(\Sigma \pm f_I)_i - 2 \cdot (\Sigma \pm f_I)_{i+1}} \cdot \frac{(\Sigma \pm f_I)_{i+1}}{(\Sigma |f_I'|)_i}, \text{where } i = 2; \\ \Delta Q_I = -\frac{(\Sigma \pm f_x)_{i+2}}{(\Sigma |f_I'|)_i \cdot \left[1 - 2 \cdot \frac{(\Sigma \pm f_I)_{i+1}}{(\Sigma \pm f_I)_i} - \left(\frac{(\Sigma \pm f_I)_{i+1}}{(\Sigma \pm f_I)_i}\right)^2\right] \cdot \left[1 - \frac{(\Sigma \pm f_I)_{i+2}}{(\Sigma \pm f_I)_{i+1}}\right] \cdot \left[1 - 2 \cdot \frac{(\Sigma \pm f_I)_{i+2}}{(\Sigma \pm f_I)_i}\right]}, \text{where } i > 2; \end{cases}$$ (13)

Formulas of flow corrections $\Delta Q_I$ depend on the counter *i*. The algorithm starts from *i* = 1, in which the multipoint method is the same as the original Newton-Raphson procedure (13a):

$$\Delta Q_I = -\frac{(\Sigma \pm f_I)_i}{(\Sigma |f_I'|)_i}$$ (13a)

In the second iteration, *i* = 2, flow corrections $\Delta Q_I$ have a little bit more complicated form (13b):

$$\Delta Q_I = -\frac{(\sum \pm f_I)_i}{(\sum \pm f_I)_i - 2\cdot(\sum \pm f_I)_{i+1}} \cdot \frac{(\sum \pm f_I)_{i+1}}{(\sum |f'_I|)_i} \tag{13b}$$

The symbol $(\sum \pm f_I)_i$ represents stored values from the first iteration, whereas $(\sum \pm f_I)_{i+1}$ represents values from the second iteration.

For the third iteration, *i* = 3, flow corrections $\Delta Q_I$ have the most complicated form (13c):

$$\Delta Q_I = -\frac{(\sum \pm f_x)_{i+2}}{(\sum |f'_I|)_i \cdot \left[1 - 2\cdot\frac{(\sum \pm f_I)_{i+1}}{(\sum \pm f_I)_i} - \left(\frac{(\sum \pm f_I)_{i+1}}{(\sum \pm f_I)_i}\right)^2\right] \cdot \left[1 - \frac{(\sum \pm f_I)_{i+2}}{(\sum \pm f_I)_{i+1}}\right] \cdot \left[1 - 2\cdot\frac{(\sum \pm f_I)_{i+2}}{(\sum \pm f_I)_i}\right]} \tag{13c}$$

This iterative process can continue, as the formula from the third iteration is used also for iterations *i*=4, 5, 6, 7 and so forth. This procedure should be done for all loops in the network separately (in our case for I, II, III, IV and V). However, in order to simplify calculations, derivative-free methods can be used [28,29].

## 4. Conclusions

Hardy Cross simplified mathematical modelling of complex problems in structural and hydraulic engineering long before the computer age. Moment distributions in indeterminate concrete structures described with differential equations were too complex for the time before computers. These finding from structural analysis, Hardy Cross later applied to balancing of flow in pipe networks. He revolutionized how the profession addressed complicated problems. Today, in engineering practice, the modified Hardy Cross method proposed by Epp and Fowler [15] is used rather than the original version of the Hardy Cross method [1]. Methods proposed by Hamam and Brameller [30], and those by Wood and Charles [31], and Wood and Rayes [32] are used in common practice [33], too. Moreover, the node oriented method proposed by Shamir and Howard [34] is also based on the Hardy Cross method.

Professional engineers use a different kind of looped pipeline in professional software [35], but even today, engineers invoke the name of Hardy Cross with awe. When petroleum and natural gas or civil engineers have to figure out what is happening in looped piping systems [36], they inevitably turn to what is generally known as the Hardy Cross method. The original Hardy Cross method is still extensively used for teaching and learning purpose [6]. Here we introduced in the Hardy Cross method the multi-point iterative approach instead of the Newton-Raphson iterative approach, but it does not affect the number of required iterations to reach the final solution in our case.

The view of Hardy Cross was that engineers lived in the real world with real problems and that it was their job to come up with answers to questions in design tasks, even if initial approximations were involved. After Hardy Cross, the essential idea which he wished to present involves no mathematical relations except the simplest arithmetic.

For example, ruptures of pipes with leakage can be detected using the Hardy Cross method because every single-point disturbance affects the general distribution of flow and pressure [37,38].

This paper has the purpose of illustrating the very beginning of modelling of gas or water pipe networks. As noted by Todini and Rossman [39], many new models have been developed since the time of Hardy Cross.

Some details about the life and work of Hardy Cross are given in Appendix B.

**Appendix A: Hydraulic models for water pipe networks and for ventilation systems**

To relate pressure $p$ [40] with flow $Q$, instead of Eq. (1), which is used for gas distribution networks in municipalities, for water distribution the Darcy-Weisbach correlation and Colebrook equation are recommended; Eq. (A.1) [23, 41-47], and for ventilation systems, the Atkinson equation; Eq. (A.2) [9]:

$$\left. \begin{array}{l} \frac{1}{\sqrt{\lambda}} = -2 \cdot log_{10}\left(\frac{2.51}{Re} \cdot \frac{1}{\sqrt{\lambda}} + \frac{\varepsilon}{3.71 \cdot D}\right) \\ \Delta p = \frac{8 \cdot \rho \cdot \lambda \cdot L \cdot Q^2}{\pi^2 \cdot D^5} \end{array} \right\} \quad (A.1)$$

$$\Delta p = \frac{\rho}{2 \cdot C_d^2 \cdot A^2} \cdot Q^2 \quad (A.2)$$

**Appendix B: The Life and work of Hardy Cross**

Hardy Cross (1885-1959) was one of America's most brilliant engineers [48-55]. He received a BSc degree in arts in 1902 and BSc degree in science in 1903, both from Hampden-Sydney College, where he taught English and Mathematics. Hardy Cross was also awarded a BSc degree in 1908 from Massachusetts Institute of Technology and an MCE degree from Harvard University in 1911, both in civil engineering. He taught civil engineering at Brown University from 1911 until 1918. He left teaching twice to get involved in the practice of structural and hydraulic engineering, from 1908 until 1909, and from 1918 until 1921. The most creative years of Hardy Cross were spent at the University of Illinois in Champaign-Urbana where he was a professor of structural engineering from 1921 until 1937. His famous article "Analysis of flow in networks of conduits or conductors" was published in 1936 in Urbana Champaign University Illinois Bulletin; Engineering Experiment Station number 286 [1]. His name is also famous in the field of structural engineering [53-55]. He developed the moment distribution method for statically indeterminate structures in 1932 [56]. This method has been superseded by more powerful procedures, but still, the moment distribution method made possible the efficient and safe design of many reinforced concrete buildings for the duration of an entire generation. Furthermore, the solution of the here discussed pipe network problems was a by-product of his explorations in structural analysis. Later, Hardy Cross was Chair of the Department of Civil Engineering at Yale, from 1937 until the early 1950s.

In 1922, Konstantin A. Čališev, emigrant from Soviet Russia, writing in Serbo-Croatian, offered a method of solving the slope deflection equations by successive approximations [57-60]. Hardy Cross improved Kališev's method as noted in [49]: "It was Hardy Cross's genius that he recognized he could bypass adjusting rotations to get to the moment balance at each and every node."

**Nomenclature**

The following symbols are used in this paper:

$\rho_r$ relative gas density (-); here $\rho_r = 0.64$

$\varrho$ density of air (kg/m³); here $\rho$=1.2 kg/m³

$L$ length of pipe (m)

$D$ diameter of pipe (m)

$Q$ flow (m³/s)

$\Delta Q$ flow correction (m³/s)

$p$ pressure (Pa)

$\Delta p$ pressure correction (Pa)

$f$ function of pressure

$f'$ first derivative of function of pressure

| | |
|---|---|
| $\lambda$ | Darcy (Moody) flow friction factor (dimensionless) |
| $Re$ | Reynolds number (dimensionless) |
| $\frac{\varepsilon}{D}$ | relative roughness of inner pipe surface (dimensionless) |
| $C_d$ | flow discharge coefficient (dimensionless) |
| $A$ | area of ventilation opening (m²) |
| $\pi$ | Ludolph number; $\pi \approx 3.1415$ |
| $i$ | counter |